\documentclass[a4paper]{ifacconf}
\usepackage{graphicx}      
\usepackage{natbib}        
\usepackage[nolist,nohyperlinks]{acronym}

\usepackage{tikz}
\usepackage{lipsum}

\newcommand\copyrighttext{%
	\footnotesize © 2025 the authors. This work has been accepted to IFAC for publication under a Creative Commons Licence CC-BY-NC-ND}
\newcommand\copyrightnotice{%
	\begin{tikzpicture}[remember picture,overlay]
		\node[anchor=south,yshift=70pt] at (current page.south) {\fbox{\parbox{\dimexpr\textwidth-\fboxsep-\fboxrule\relax}{\copyrighttext}}};
	\end{tikzpicture}%
}

\begin{document}

\begin{acronym}
	\acro{API}{Application Programming Interface}
	\acro{AAS}{Asset Administraion Shell}
	\acro{PnID}[P\&ID]{Piping \& Instrumentation Diagram}
	\acro{FEED}{Front End Engineering \& Design}
	\acro{BFD}{Block Flow Diagram}
	\acro{DCS}{Distributed Control Systems}
	\acro{EPC}{Engineering, Procurement, and Construction}
	\acro{PLC}{Programmable Logic Controller}
	\acro{DT}{Digital Twin}
	\acro{IDTA}{Industrial Digital Twin Association}
	\acro{MTP}{Module Type Package}
	\acro{HMI}{Human Machine Interface}
	\acro{PEA}{Process Equipment Assembly}
	\acro{POL}{Process Orchestration Layer}
	\acro{FM}{Function Module}
	\acro{OPC UA}{Open Platform Communication Unified Architecture}
	\acro{IoT}{Internet of Things}
	\acro{AID}{Asset Interfaces Description}
	\acro{VCS}{Version Control System}
	\acro{BPMN}{Business Process Model and Notation}
	\acro{RBAC}{Role-Based Acessed Control}
	\acro{ID}{Identifier}
	\acro{SMC}{Submodel Element Collection}
	\acro{DEXPI}{Data Exchange in the Process Industry}
\end{acronym}

\begin{frontmatter}

\title{Towards an Engineering Workflow Management System for Asset Administration Shells using BPMN}

\copyrightnotice

	

\author[ABB]{Sten Grüner} 
\author[ABBCR]{Nafise Eskandani} 

\address[ABB]{Process Control Platform, ABB AG, 
	Mannheim, 68309 Germany (e-mail: sten.gruener@de.abb.com).}
\address[ABBCR]{ABB AG Corporate Research Center Germany, 
   Mannheim, 68309 Germany (e-mail: nafise.eskandani@de.abb.com).}

\begin{abstract}                
The integration of Industry 4.0 technologies into engineering workflows is an essential step toward automating and optimizing plant and process engineering processes. The \ac{AAS} serves as a key enabler for creating interoperable Digital Twins that facilitate engineering data exchange and automation. This paper explores the use of AAS within engineering workflows, particularly in combination with \ac{BPMN} to define structured and automated processes. We propose a distributed AAS copy-on-write infrastructure that enhances security and scalability while enabling seamless cross organizational collaboration. We also introduce a workflow management prototype automating AAS operations and engineering workflows, improving efficiency and traceability. 
Our results indicate a great potential of BPMN on both AAS infrastructure and business process sides.
\end{abstract}

\begin{keyword}
Process and Plant Engineering Workflow, Automation, Industry 4.0, Digital Twin, Asset Administration Shell, AAS, Business Process Model and Notation, BPMN 
\end{keyword}

\end{frontmatter}

\section{Introduction}
As industries digitize, the need for standardized, interoperable, and scalable solutions for engineering processes has become more evident. 
One of the fundamental pillars supporting this transformation is the concept of the Asset Administration Shell (\ac{AAS}). 
Initially developed as part of Germany’s Industry 4.0 initiative, AAS provides a structured digital representation of assets, ensuring seamless data exchange and interaction between different organizations and systems. 

Traditional engineering processes often rely on scattered, non-standardized data exchanges, leading to inefficiencies, delays, and increased costs. The absence of structured workflows hinders consistency and traceability across project phases. When integrated into engineering workflows, \ac{AAS} introduces a structured approach through predefined submodels that encapsulate asset-related information in a machine-readable format to enhance collaboration and ensure consistency throughout an asset's lifecycle. However, AAS alone does not fully address fragmented and inefficient workflows. To maximize its potential, well-defined and automated processes are needed to regulate data flow, decision-making, and change management. This is where Business Process Model and Notation (\ac{BPMN}) plays a crucial role by providing a formalized framework for defining, executing, and monitoring workflows. By integrating BPMN with AAS, engineering tasks can be systematically orchestrated to ensure traceability, compliance, and resource optimization.

Implementing a distributed AAS infrastructure and integrating BPMN into engineering workflows pose significant challenges. Organizations operate within complex ecosystems with diverse stakeholders, systems, protocols, and security requirements. Ensuring secure, scalable data exchange requires an interoperable yet tightly controlled infrastructure. Additionally, workflow automation must be flexible enough to accommodate various industrial use cases while maintaining necessary human oversight. Overcoming these challenges is crucial to fully leveraging AAS and BPMN in industrial applications.

This paper addresses the challenge of integrating AAS into engineering workflows by proposing a distributed AAS infrastructure that enhances security, scalability, and cross organizational collaboration. We present a workflow management system that leverages BPMN to automate AAS operations, improving efficiency and traceability. Through a prototype implementation, we demonstrate how such a system can facilitate engineering processes, enabling organizations to achieve greater consistency and interoperability in industrial automation.

The rest of this paper is structured as follows: Section~\ref{sec:intro2} introduces \ac{AAS} and BPMN. 
Section~\ref{sec:eng} reviews the usage of \ac{AAS} in process and plant engineering.
Section~\ref{sec:cow} introduces a novel copy-on-write approach to operate distributed AAS infrastructure. 
Section~\ref{sec:bpmn} defines a set of requirements for using workflow management systems for AAS infrastructure as well as for business process described with AAS.
Section~\ref{sec:proto} reviews the developed workflow management prototype.
Finally, a summary and an outlook is provided after a short review or related work in Section~\ref{sec:realted}.

\section{AAS and BPMN}
\label{sec:intro2}
The concept of \ac{AAS} is one of the corner stone contributions of German Industry 4.0 community which has become an IEC (IEC 63278) standard for creating interoperable Digital Twins in industrial applications.
\ac{AAS} is a digital representation of a physical or digital asset which structures asset information (e.g., properties, events, or operations) in submodels (i.e., standalone information containers).
The information model of an \ac{AAS} can be serialized into multiple formats, e.g., XML or JSON, and be transferred as a file (so-called AASX-package for a type 1 AAS) or via a reactive RESTful API with JSON payload (so-called type 2 AAS).

Core specifications of the AAS focus on the metamodel and technical aspects of \ac{AAS} like API definitions. The main step towards standardization and interoperability of the content of the AAS are a set of so-called submodel templates -- blueprints for information payload of an AAS for a specific use-case. The development of submodel templates is governed by \ac{IDTA}, a non-profit association having more than 100 international members from different sectors like software, industrial equipment vendors, energy and automotive industry, as well as academic partners.
Currently there are around 90 submodel templates which are published or under the standardization covering use cases like digital nameplate, handover documentation, or asset location.

Some IDTA-standardized AAS submodel templates are particularly relevant for engineering in process industries. Examples are standardized submodel temples for \ac{DEXPI} (cf.~\cite{wiedau2019enpro}) standard for \acp{PnID} or \ac{MTP} standard for modular automation (cf.~\cite{tauchnitz2022mtp}).

\ac{BPMN}, cf.~\cite{chinosi2012bpmn}, is a mature standardized (ISO 19510) graphical notation for business processes (i.e., a collection of steps performed by employees or equipment to produce a product or a service). 
\ac{BPMN} allows to visually depict workflows with the help of flowchart-like elements like decision gateways, events, swim lanes, and activities. An example of \ac{BPMN} notation can be found in Fig.~\ref{fig:bpmn}.

Wide usage of BPMN across various industries spans documentation, optimization, as well as automation of business processes use cases. 
In the latter case, an execution engine is needed to execute the business process defined with \ac{BPMN}. An engine is typically an enterprise software executing BPMN processes following the standardized semantics. This typically involves orchestrating and visualizing the execution of steps within the process, e.g., sending an email or event, offering a human user to perform some actions, or communicating with an external system via an \ac{API}.

Being typical enterprise software, BPMN engines offer a wide range of features, including user management, user action tracing, and reporting. However, they tend to prioritize these enterprise functionalities over real-time capabilities and efficient computing resource utilization.

\section{AAS-based Continuous Engineering}
\label{sec:eng}
\begin{figure}
	\begin{center}
		\includegraphics[width=0.98 \columnwidth]{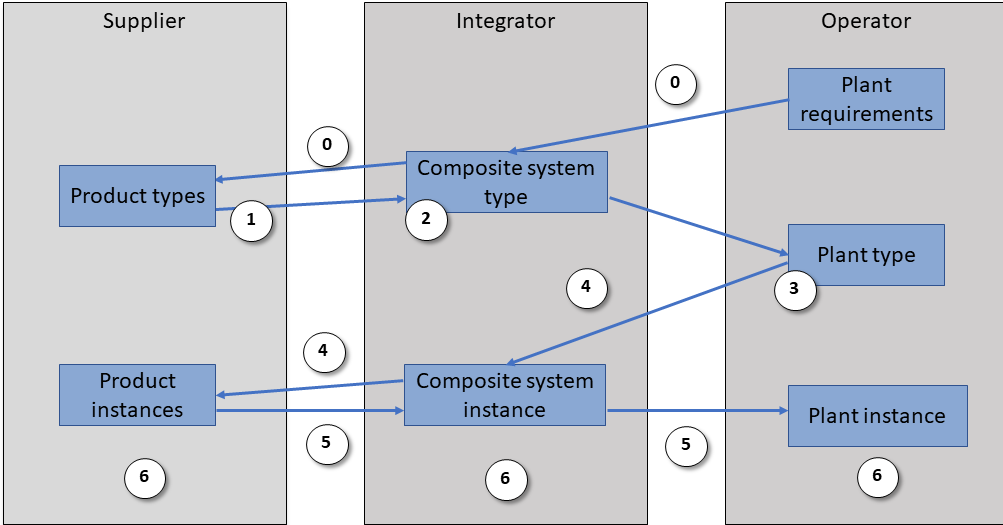}    
		\caption{Cross organizational information exchange (figure from \cite{GrünerHoernickeStarkSchochEskandaniPretlove+2023+689+708})} 
		\label{fig:participants}
	\end{center}
\end{figure}

Approaches to applying Industry 4.0 technology stack (including \ac{AAS}) in plant engineering have been discussed by the community since 2017 (\cite{wagner2017role}).
\cite{GrünerHoernickeStarkSchochEskandaniPretlove+2023+689+708} introduced an AAS-based engineering workflow designed for collaboration between multiple organizations. This workflow builds upon the information exchange frameworks established by the German ``Plattform Industrie 4.0'' initiative and has been adapted to meet the specific requirements of the energy and oil and gas industries, as illustrated in Fig.~\ref{fig:participants}.

In a nutshell, the engineering process starts with the plant requirements (bubble 0) which are handed over from plant owner/operator (e.g., an energy company) to the integrator (or \ac{EPC}). Based on the requirements, \ac{EPC} creates a blueprint of a composite system (bubble 2). In general, this systems might be a whole plant or a smaller plant segment, e.g., an oil-water separator with multiple separation stages. 
Furthermore, many of those composite systems are instantiated several times, e.g., the above mentioned separator can be used multiple times in the context of the same plant or different installations of the same operator.
In order to create the blueprint, the \ac{EPC} needs to select equipment of suppliers, e.g., sensors or actuators, based on operator's requirements (bubble 0). Supplier provides catalog information for devices to the \ac{EPC} (bubble 1).

After several iterations, the plant type is approved by the operator (bubble 3) and the ordering/production of the actual plant and utilized physical equipment may begin (bubbles 4 and 5) in an emergent physical plant. Based on the digital information from the engineering process, lifetime services can be offered between all three parties (bubble 6), e.g., predictive maintenance of supplier's devices or plant upgrades provided by \acp{EPC}. 

Along with the plant design and engineering workflow, different artifacts are exchanged between organizations (i.e., a plant owner/operator, an \ac{EPC}, and an equipment manufacturer). Currently those artifacts are file-based and often either not machine-readable (e.g., PDF documents like \acp{PnID}) or lack standardized machine-readable semantics (e.g., spreadsheet-based signal lists). 

In \cite{GrünerHoernickeStarkSchochEskandaniPretlove+2023+689+708}, we proposed the AAS as a modeling and implementation backbone digitizing existing engineering workflows between parties. A set of IDTA submodel templates was identified, which partially covered the relevant use cases in the energy industry.

Still, in order to ``compete'' with established proprietary engineering systems within the domain, several missing features were identified. These include (1) versioning of the AAS and its associated information (such as submodels and their components, including attached engineering documents) to enable change traceability, and (2) best practices for operating a distributed AAS infrastructure that supports distributed access and cross organizational, event-driven communication.

Subsequent work (\cite{gruner2024asset}) described results of a collaborative study between ABB and a Norwegian energy company Equinor where the above mentioned workflows were realized based on open-source (Eclipse BaSyx \cite{10.1145/3459960.3459978}) implementations of \ac{AAS} infrastructure. From the perspective of identified feature gaps, following solutions were proposed.

Wrt.\ the first gap, versioning and marking checkpoints of information within the \acp{AAS} was prototypically solved by connecting AAS infrastructure to git \ac{VCS}. Technically, a snapshot of server's content was created based on the XML-based serialization of the \ac{AAS} including the attached files (as AASX package). 
This allowed tagged \ac{VCS} commits (e.g., on certain points like quality gates) and fine-grained traces of changes within AAS based on the differences of XML files. 
	
This approach suits for information that changes infrequently, as is common in engineering workflows. Adopting git-flow-based workflows was also feasible, leveraging well-established version control practices from platforms such as GitHub and GitLab (e.g., pull requests). 

\section{Copy-on-write AAS Infrastructure Operation Mode}
\label{sec:cow}
Regarding the second gap, a \emph{copy-on-write} operation mode for distributed \ac{AAS} infrastructure was proposed. Here, the generic agreement between the participating organizations is made, s.t., \ac{AAS} modifications is only possible by the respective organization hosting the servers while cross organizational read-access is generally possible based on the need-to-know principle.

\begin{figure}
	\begin{center}
		\includegraphics[width=0.9\columnwidth]{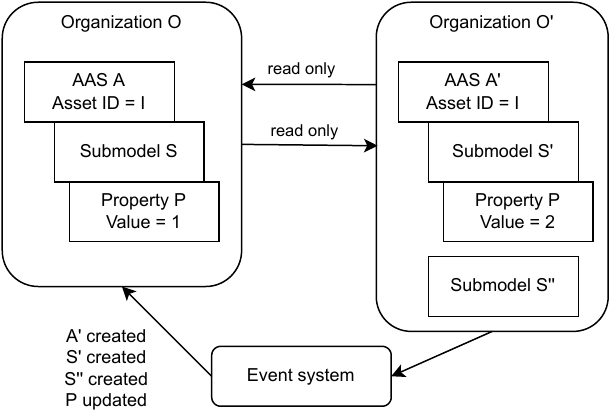}    
		\caption{Copy-on-write operation mode for distributed AAS infrastructure} 
		\label{fig:cow}
	\end{center}
\end{figure}
An example of this principle is presented in Fig.~\ref{fig:cow}: Organization $O$ hosts the AAS $A$ with some information in submodel $S$. Organization $O'$ has only a read access to this information. 
In case a change of information is required by $O'$ in $A$, $O'$ creates a new \ac{AAS} $A'$ shearing the same asset \ac{ID} (named $I$) of $A$. 
If change is needed within the submodel $S$ of $A$, a copy $S'$ of $S$ is linked to $A'$ and updated, i.e., a property $P$ is modified. 
Addition of new submodels is also possible by adding a new submodel $S''$ to $A'$ which was originally not present in the \ac{AAS} $A$.

General read rights across organizations enables querying all known AAS discovery endpoints (to resolve an asset \ac{ID} to a set of \acp{AAS} referring to it). These queries allow each party to discover all information relevant to the asset, i.e., organization $O$ is able to detect an additional \ac{AAS} $A'$ hosted by $O'$ with all changed and added information.

Furthermore, with a suitable event mechanism, e.g., via an event broker, each party can receive updates on changed and added \ac{AAS}s and submodels and react to the change accordingly. Eclipse BaSyx provides a non-standardized implementation of such MQTT-based event mechanisms used in our prototype.

The read only restriction can be technically implemented by using \ac{RBAC} features of Eclipse BaSyx or a web-application firewall restricting access to some verbs of the RESTful API of the \ac{AAS} (e.g., blocking POST and PUT requests). 

The presented operation principle has several advantages for use-cases where the \ac{AAS} and submodel origin can be clearly attributed to one organization:
\begin{itemize}
	\item No uncontrollable/unwanted modifications on information not owned by the particular organization.
	\item High level of redundancy through local information storage lowers the availability requirements the infrastructure of each organization, i.e., the engineer process is not blocked in case of server unavailability for some organization.
\end{itemize}

These advantages come at a cost of following restrictions:
\begin{itemize}
	\item Duplication of data requires careful versioning and version tracking of \acp{AAS} and submodels.
	\item Duplication might get unpractical in scenarios where information is changed at high frequency, e.g., \ac{IoT} or telemetry data.
	\item \ac{AAS} SDKs and tools need to cope with multiple endpoint sets for each organization requiring different authentication information sets and access tokens. 
\end{itemize}

\begin{figure}[t]
	\begin{center}
		\includegraphics[width=0.8\columnwidth]{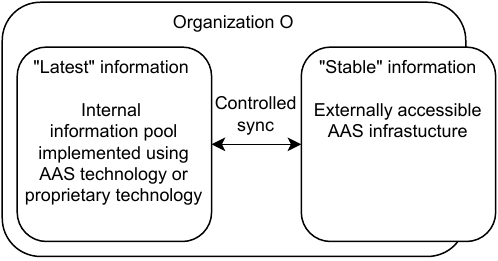}    
		\caption{``Latest'' and ``stable'' information sets} 
		\label{fig:latest}
	\end{center}
\end{figure}
Despite the disadvantages, we believe that the proposed copy-on-write infrastructure operation mode is well-suited for applications in engineering where the information is not changed on a high rate. This is due to the fact that changes are published in larger change-sets, e.g., before reaching a certain quality gate or project delivery stage.

These two information sets are seen in Fig.~\ref{fig:latest}. The ``stable'' set, accessible via externally available AAS endpoints, represents a collection of AAS instances and submodels (such as those for various sensors in a plant) to ensure that data remains in a consistent and validated state.
On the other hand, there is a ``latest" information set representing a ``live view" into the engineering tools of the respective organization. The need for the process of controlled sync was also mentioned already by \cite{Drath2024-rl}.

Thus, Fig.~\ref{fig:cow} shows only externally visible, ``stable'' endpoints serving a released, consistent, and quality-approved state of engineered information via \ac{AAS} and its submodels. Intermediate or ``latest'' changes reside in non-AAS storage (e.g., legacy engineering tools) or a ``staging'' AAS infrastructure accessible only within one organization.

\section{Workflow management for \ac{AAS}}
\label{sec:bpmn}
\subsection{Workflows for \ac{AAS} infrastructure}
\label{sec:bpmn4infra}
With the progress of the collaborative study, the amount of used \acp{AAS} grew to several dozens and the so far manually triggered infrastructural operations like cloning (creating \ac{AAS} $A'$ based on $A$ while preserving the asset \ac{ID} $I$) became too tedious and error-prone.
A possible solution to this infrastructural problem is a workflow management system fulfilling the following requirements:
\begin{itemize}
	\item Workflows need to be instantiated out of pre-created workflow templates based on certain infrastructure conditions (e.g., upon an availability of a new AAS in the \ac{AAS} repository of an organization).
	\item Workflows and execution of workflow instances can be graphically defined and monitored, respectively.
	\item Workflows events guarding transitions between tasks can be triggered by AAS infrastructure.
	\item Workflows allow automatic task execution and manual steps for users (e.g., alternative selection).
	\item Worflow engine defines multiple user groups based on performed tasks (e.g., process and plant engineers).
	\item The workflow instance execution and especially the tasks of users are traceable for eventual audits.
\end{itemize}

\ac{BPMN} was chosen as a candidate for workflow formalization based on our positive prior experiences with the language in the context of Industry 4.0, where it has been used for equipment orchestration (e.g., in \cite{9921694}). We evaluate its alignment with the requirements through the prototype developed, presented in Section~\ref{sec:proto}. 

\subsection{AAS Submodel Workflows Across Organizations}

Despite the focus of this work on the AAS infrastructure, we see a great potential in applying workflows (and especially \ac{BPMN} in its home turf) for cross organizational business processes where entities are represented as assets, respective \ac{AAS} and \ac{AAS} submodels.

An example is a service request notification submodel template released by \cite{IDTA02010-1-0}. This submodel template defines a set of building blocks to store all the information needed to issue request for service for the asset (compared with a ``call for help'' button in the submodel template specification).
An instance of the service request submodel consists of a set of service request notifications contain, among others, the information about the organization requesting the service (such as customer identity and general contact information).
Additionally submodel elements for detailed information about the fault with the possibility to attach media, e.g., photos or screenshots, are provided.
Finally, information for on site contact is made available.

The submodel template specifies the ``data set'' for each service request quite thoughtfully by providing semantic \acp{ID} for each element and defining interoperable enumerations (e.g., type of requested services) based on ECLASS \acp{ID}.
However, the process that follows the creation of a service request as an \ac{SMC} within the respective submodel is not clearly defined. According to the submodel template’s textual description, either the server hosting the submodel can communicate with a remote service system (e.g., that of the asset vendor), or a local asset management system within the requester's organization can detect the newly added service request and initiate follow-up actions.

\begin{figure}
	\begin{center}
		\includegraphics[width=1\columnwidth]{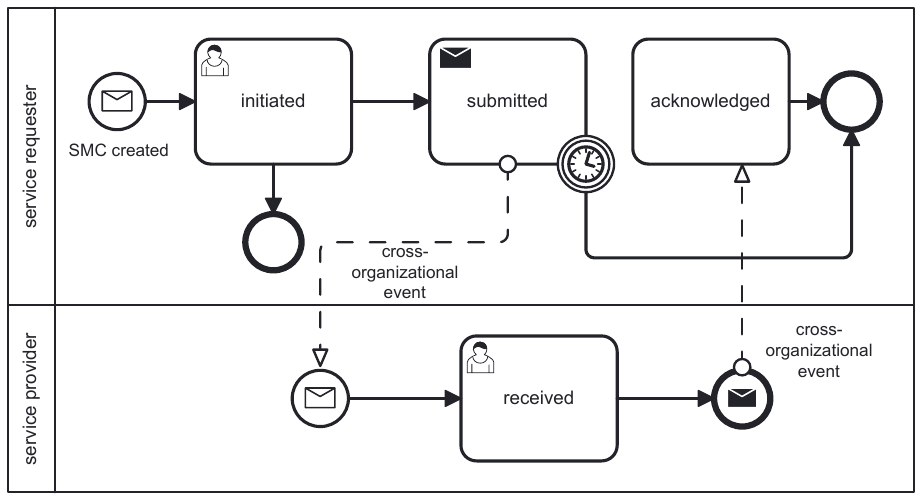}    
		\caption{A mocked-up BPMN representation for a cross organizational business workflow realized with AAS} 
		\label{fig:bpmn}
	\end{center}
\end{figure}

Integrating BPMN workflows and cross-AAS events enables further standardization of this process to ensure a more structured and interoperable approach. A conceptual representation of this workflow is illustrated in Fig.~\ref{fig:bpmn}, where the following sequence of operations takes place:
\begin{enumerate}
	\item Creation of the service request \ac{SMC} within the submodel, creates a workflow instance based on \emph{standardized} predefined workflow template (e.g., as a potential appendix to the submodel template).
	\item A next example step can be a task for the user, e.g., to confirm and initiate the service request.
	\item If initiated, the process emits a cross organizational AAS event towards service provider AAS infrastructure. Following the description within the submodel template, the server serving the AAS can emit this event or it can be delegated to some asset management system within the requester's organization.
	\item The workflow goes into a ``submitted'' state, which can be tracked for each service requests. Additionally, a time out is defined to terminate the workflow in case no confirmation was received.
	\item Later, the current state can change to some additional state, e.g., ``acknowledged'', based on events received from the service provider, e.g., after a manual receive confirmation step on provider's side. 
\end{enumerate}

A standardized set of steps would allow to further advance digitalization with help of Industry 4.0 technology.

\section{Workflow Management Prototype}
\label{sec:proto}

In this section we present our prototype developed for the AAS infrastructure use case described in Section~\ref{sec:bpmn4infra}.

\begin{figure}
	\begin{center}
		\includegraphics[width=1\columnwidth]{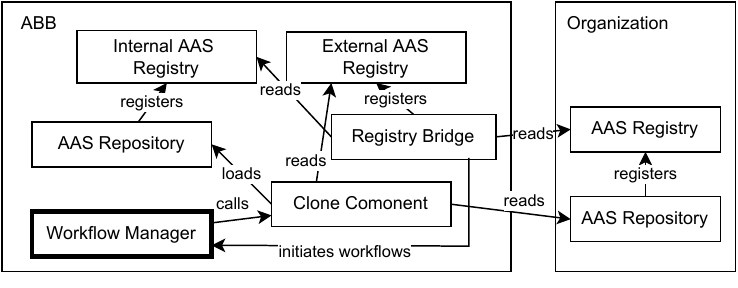}    
		\caption{Architecture of our AAS infrastructure combined with a workflow manager (simplified)} 
		\label{fig:arch}
	\end{center}
\end{figure}

Our system is based on the AAS component architecture which was discussed in detail in previous works (\cite{gruner2024asset}). In this work we focus on key components depicted in Fig.~\ref{fig:arch}.
For reasons of comprehensiveness, we omit a number of key components in Fig.~\ref{fig:arch}, e.g., user-facing UI and APIs delivered through a reverse proxy, or an MQTT message broker connected to both organizations. Furthermore, submodel registries and submodel repositories are omitted from the figure. In principle, they work in the same way as AAS repositories and registries.

In the following, we provide an overview of the key components deployed as software-container-based microservices:
\begin{itemize}
	\item AAS Repositories of both organizations host \acp{AAS} and provide references to submodels hosted by different servers (omitted in the figure).
	\item AAS Repositories provide an API to look up all available \acp{AAS} of the respective organization.
	\item ABB maintains an 'Internal' AAS Registry listing its own AASs. Additionally, an 'External' AAS Registry is available, which includes both ABB's AASs and those of external organizations.
	\item Registry Bridge is a  dedicated component which synchronizes the External AAS Registry with information from registries of both organizations. Synchronization is done interval-triggered or event-triggered through MQTT broker (omitted in figure).
	\item Clone Component copies \acp{AAS} from the infrastructure of Organization into ABB's infrastructure as described in Section~\ref{sec:bpmn4infra}. It changes IDs of AAS and submodels while preserving asset IDs.
\end{itemize}

The introduced Workflow Manger component demonstrates how the described cloning procedure can be performed according to the requirements elicited in Section~\ref{sec:bpmn4infra}. 
Due to some unclarity with the licensing conditions of Camunda BPMN engine, we selected another Java-based BPMN engine called Activiti which was deployed as software container along with the rest of the components.

\begin{figure}
	\begin{center}
		\includegraphics[width=1\columnwidth]{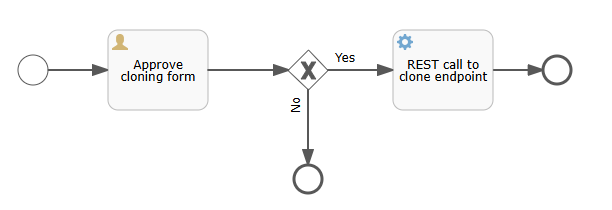}    
		\caption{AAS clone process template} 
		\label{fig:template}
	\end{center}
\end{figure}

The simplest process template selected for the example is shown in Fig.~\ref{fig:template}.
Once a process instance is created, a user task ``Approve cloning form" allows user has to make a decision whether a new AAS has to be cloned or not. 
For this task, a user form is provided, containing additional information like identification of the AAS. Based on the logged decision, the decision gate either terminates the process instance or goes to the next task where the Clone Component is invoked using its REST API.

\begin{figure}
	\begin{center}
		\includegraphics[width=0.8\columnwidth]{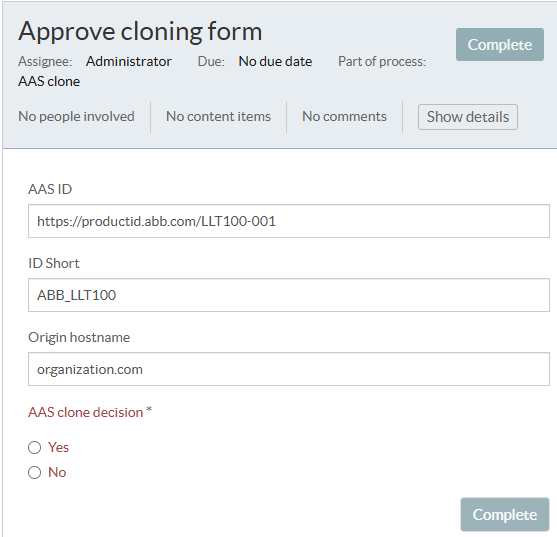}    
		\caption{Task form as screenshot from Activiti UI} 
		\label{fig:form}
	\end{center}
\end{figure}

For the selected use case of AAS cloning, the clone processes within the Workflow Manager are initiated by the Registry Bridge in case a new AAS or submodel are detected in Organization's infrastructure (e.g., based on MQTT events). 
Initiation is done via an API call of the Activiti engine, s.t., a clone workflow instance is created for every new AAS. 
Functionality of Activiti workflow engine allows then to track the state of each of those processes, i.e., decisions made by the user.
Activiti also supports other requirements listed in Section~\ref{sec:bpmn4infra} since those features are typically basic for workflow engines.

\section{Related Work}
\label{sec:realted}

Challenges for AAS modeling and infrastructure used in context of plant engineering have been already identified in previous work \cite{GrünerHoernickeStarkSchochEskandaniPretlove+2023+689+708}. \cite{Drath2024-rl} continued to analyze the engineering requirements for \acp{AAS} and identify, among others, two points of criticism wrt.\ to missing ``commit'' protocol for released and intermediate engineering data as wells as standardized workflows which are addressed in this contribution.

From the perspective of plant and automaton engineering, a set of structured and standardized workflows for engineering are missing probably because of complex customer-specific or even project-specific customization.
One possible standardization candidate known to us is the NAMUR worksheet NA 35 (cf.~\cite{na35}). It defines basic foundations of process automation projects with larger activities like basic and detailed engineering, as well fine-grained breakdowns of those.

BPMN was utilized to define a support system for process and automation engineering in \cite{bigvand2017workflow}, however no cross organizational information information flow was considered in that work. Furthermore, BPMN was also applied in the context of controlling production flows based on \ac{AAS} interfaces in \cite{9921694}. In our approach we apply BPMN workflows to orchestrate AAS infrastructure as well as project its usage to pure business workflows implemented using AAS.

Apart from BPMN, there is a number of contributions to define continuous data flows between \acp{AAS} for typical IoT or telemetry applications. Examples are applications of Node-RED in \cite{vaage2023industry} or custom BaSyx components in \cite{10275536}.

Additionally, work around so-called type 3 or proactive \ac{AAS} is highly relevant in context of this paper. Type 3 AAS can directly interact to other type \acp{AAS} based on Industry 4.0 language specified in German specification VDI/VDE 2193 as summarized in \cite{grunau2022implementation}.

Based on the concepts of type 3 AAS, an implementation of a BPMN-based bidding workflow for automated production planning executed via a workflow engine was presented in \cite{10710731}. In fact this is a much more sophisticated and mature implementation compared to the envisioned use of BPMN processes we project in Fig.~\ref{fig:bpmn}. However, our work is covering use cases of standardized workflows within submodels, operating BPMN workflows directly on the elements of AAS (such as \ac{SMC}), as well as interacting with AAS infrastructure for cloning which were not covered in \cite{10710731}.

\section{Summary and Future Work}
This paper has outlined a distributed AAS infrastructure that facilitates secure cross organizational collaboration and has demonstrated a prototype workflow management system that automates AAS operations. The integration of AAS with BPMN-based workflow management systems represents a significant advancement in engineering automation. By leveraging AAS for structuring asset information and BPMN for defining workflows, organizations can achieve greater efficiency, traceability, and interoperability in industrial engineering processes. 

Future work includes refining AAS-based workflows by incorporating AI-driven optimization techniques, further standardizing submodel templates for diverse industrial applications, and enhancing real-time capabilities for high-frequency data updates. The adoption of AAS-based engineering workflows has the potential to extend beyond industrial automation, finding applications in fields such as smart manufacturing, supply chain management, and predictive maintenance. By continuing to develop and standardize AAS workflows, the industry can move closer to fully automated, data-driven engineering ecosystems that optimize efficiency and reduce operational costs.


\bibliography{ifacconf}             
                                                   







\end{document}